\documentclass[twocolumn]{article}
\usepackage{fullpage}
\usepackage[pdftex]{graphicx}
\usepackage{times}
\usepackage{fancyvrb}
\usepackage{tabularx}

\newcommand{\textsec}{{\tt .text} section}
\newcommand{\textsecs}{{\tt .text} sections}
\newcommand{\comsec}{{\tt .comment} section}

\newif\ifremark
\long\def\remark#1{
\ifremark%
    \begingroup%
    \dimen0=\textwidth
    \advance\dimen0 by -1in%
    \setbox0=\hbox{\parbox[b]{\dimen0}{\protect\em #1}}
    \dimen1=\ht0\advance\dimen1 by 2pt%
    \dimen2=\dp0\advance\dimen2 by 2pt%
    \vskip 0.25pt%
    \hbox to \textwidth{%
        \vrule height\dimen1 width 3pt depth\dimen2%
        \hss\copy0\hss%
        \vrule height\dimen1 width 3pt depth\dimen2%
    }%
    \endgroup%
\fi} \remarktrue
\begin{document}

\title{
{\bf Automatically Mining Program Build Information via Signature Matching}
}

\author{
\begin{tabular}{c}
Charng-Da Lu\\
Buffalo, NY 14203
\end{tabular}
}
\date{}

\maketitle
\begin{abstract} Program build information, such as compilers and libraries
used, is vitally important in an auditing and benchmarking framework for HPC systems.
We have developed a tool to automatically extract this information using
signature-based detection, a common strategy employed by anti-virus software to
search for known patterns of data within the program binaries. We formulate the
patterns from various ''features'' embedded in the program binaries, and the
experiment shows that our tool can successfully identify many different
compilers, libraries, and their versions.
\end{abstract}

\section{Introduction} One important component in an auditing and benchmarking
framework for HPC systems is to be able to report the build information of
program binaries. This is because the program performance depends heavily on
the compilers, numerical libraries, and communication libraries. For example,
the SPEC CPU 2000 Run and Reporting Rules \cite{SPEC} contain meticulous
guidelines on the reporting of the compiler of choice, compilation flags,
allowed and forbidden compiler tuning, libraries, data type sizes, etc.

However, in most HPC systems, program build information, if maintained at all,
is recorded manually by system administrators. Over time, the sheer number of
software/library packages of different versions, builds, and compilers of
choice can grow exponentially and become too daunting and burdensome to
document. For example, at our local center we have software packages built from 250
combinations of different compilers and numerical/MPI libraries. On
larger systems such as Jaguar and Kraken at the Oak
Ridge National Laboratory, the number can be as high as 738 \cite{ALTD}.

In addition, there is no standard format of documenting program build
information. Many HPC systems use Modules \cite{Modules} or SoftEnv
\cite{SoftEnv} to manage software packages, and a common naming scheme is to
incorporate the compiler name (as a suffix) in the package name. There is
usually additional textual description to indicate build information, such as
compiler version, debug/optimization/profiling build, and so on. Mining these
free-form texts, however, requires the understanding of each HPC site's software
environment and documentation style and is not generally applicable.

In this paper, we present a signature-matching approach to automatically
uncover the program build information. This approach is akin to the
common strategy employed by anti-virus software to detect malware: search
for a set of known signatures. We exploit the following "features'' of
program binaries and create signatures out of them:
\begin{itemize}\itemsep0pt
\item Compiler-specific code snippets.
\item Compiler-specific meta data.
\item Library code snippets.
\item Symbol versioning.
\item Checksums.
\end{itemize}

Our approach has several advantages. First, we only need to create, annotate,
and maintain a database of signatures gathered from compilers and
libraries, and we can then run the signature scanner over program binaries to
derive their build information. Second, unlike the anti-virus industry where
the malware code must be identified and extracted by experts, our signature
collection process is almost mechanical and can be performed by non-experts.
Third, our approach does not rely on symbolic information and thus can handle
stripped program binaries.

Our implementation is based on the advanced pattern matching engine
of ClamAV \cite{ClamAV}, an open-source anti-virus package. We choose ClamAV
for its open-source nature, signature expressiveness and scanning speed.

The remainder of this paper begins by describing the features in the program
binaries. Section~\ref{sec:Implementation}-\ref{sec:Evaluation} provide
the implementation details and experimental results. We then discuss
potential improvement and related work in \S~\ref{sec:Discussion}-\ref{sec:RelatedWork},
followed by a conclusion in \S\ref{sec:Conclusion}.

\section{Program Binary Characteristics}
On most modern UNIX and UNIX-related systems, the executable binaries (programs
and libraries) are stored in a standard object file format called the
Executable and Linking Format (ELF) \cite{ELF,ELF2}. An ELF file can be
divided into named "sections,'' each of which serves a specific function at
compile time or runtime. The sections relevant to our work are:
\begin{itemize}\itemsep0pt
\item \textsec{} contains the executable machine code and is the
 main source for our signature identification.
%
%
\item \comsec{} contains compiler and linker specific version control
information. More on this in \S\ref{subsec:CompilerSpecificMetaData}.
\item {\tt .dynamic} section holds dynamic linking information, including
file names of dependent dynamic libraries, and pointers to symbol version
tables and relocation tables.
\item {\tt .rel.text} and {\tt.rela.text} sections consist of relocation
tables associated with the corresponding \textsecs{}. More details in
\S\ref{subsec:SignatureGenerator}.
\item {\tt .gnu.version\_d} section comprises the version definition table.
More on this in \S\ref{subsec:SymbolVersioning}.
\end{itemize}

There is a wealth of information embedded in these sections, and in the following
we explain these characteristics in detail.

\subsection{Compiler-Specific Code Snippets}
\label{subsec:CompilerSpecificCodeSnippets}
It is not news that certain popular compilers on the Intel x86
platform insert extra code snippets unbeknownst to the developers
\cite{Agner}. We will illustrate with three examples.

The first example is the so-called "processor dispatch" employed by certain
optimizing compilers. As the x86 architecture evolves with the addition of new
capabilities and new instructions such as Streaming SIMD Extensions (SSE) and
Advanced Vector eXtensions (AVX), an optimizing compiler will produce machine
code tuned for each capability.  Since the new instructions are not
recognized by older generations of x86 processors, to avoid "illegal
instruction'' errors and to re-route the execution path to the suitable code
blocks, an extra code snippet is inserted to perform this task.

Both Intel and PGI compilers, when invoked with optimization flags enabled
(and {\tt -O2} is used implicitly), insert the processor dispatch
code which is executed before the application's {\tt main} function. These code snippets
invariably use the {\tt cpuid} instruction to obtain processor feature flags.
For example, the core processor dispatch routine used by the Intel compiler
is called {\tt \_\_intel\_cpu\_indicator\_init}. It initializes an internal variable
called {\tt \_\_intel\_cpu\_indicator} to different values based on the processor on
which the program is running \cite{Agner}. This information is later used to either abort
program execution immediately, with an error like "This program was
not built to run on the processor in your system,'' or execute different code
blocks (tuned for different generations of SSE instructions) in Intel's
optimized C library routines such as {\tt memcpy} and {\tt strcmp}.

A second instance of compiler-inserted code is to enable or disable certain floating-point
unit (FPU) features. For example, when GCC is invoked with {\tt -ffast-math} or
{\tt -funsafe-math-optimizations} optimization flags, it inserts code to turn on
the Flush-To-Zero (FTZ) mode and the Denormals-Are-Zero (DAZ) mode in the x86
control register {\tt MXCSR}. When these modes are on, the FPU bypasses IEEE 754
standards and treats denormal numbers, i.e. values extremely close to zero, as
zeros. This optimization trades off accuracy for speed \cite{Denormal}.
The GNU C Compiler, GCC, also accepts {\tt -mpc\{32|64|80\}} flags,
which are used to set the legacy x87 FPU precision/rounding mode.
Again, GCC uses a special prolog code to configure the FPU to the requested mode.

A third instance of compiler-inserted code is to initialize user's data.
For example, one of the C++ language features requires that static objects must be
initialized, i.e. their constructors must be called, before program
startup \cite{CppStandard}. To implement this, the C++ compiler emits a
special ELF section called {\tt .ctors}, which is an array of pointers to static
objects' constructors, and inserts a prolog code snippet which
sweeps through the {\tt .ctors} section before running the application's {\tt main} function.

\subsection{Compiler-Specific Meta Data}
\label{subsec:CompilerSpecificMetaData}
ELF files have an optional section called {\tt .comment} which consists of a
sequence of null-terminated ASCII strings. This section is not loaded into
memory during execution and its primary use is a placeholder for version control
software such as CVS or SVN to store control keyword information. In practice,
most compilers we examined will also fill this section with strings which are
unique enough to differentiate the compilers and the versions
(see \S\ref{subsec:CompilerIdentification}). The compiler adds string data by
using the {\tt .ident} assembler directive when generating the assembly code,
and then the assembler pools these strings and saves them into the \comsec{}.
Unlike the debugging and symbolic information embedded in other ELF sections,
the \comsec{} is not removed by the GNU {\tt strip} utility, so we can mine it to
obtain the compiler provenance.

For example, using the GNU {\tt readelf} tool with command-line option
{\tt -p .comment} on GCC-compiled programs could have the following output:
{\small
\begin{Verbatim}
GCC: (GNU) 4.1.2 20080704 (Red Hat 4.1.2-50)
\end{Verbatim}
}

\subsection{Library Code Snippets}
\label{subsec:LibraryCodeSnippets}
If a program calls library functions, the linker will bind the functions to
libraries to create the executable. The linking mode is either static or
dynamic. In the former, the linker extracts the code of called functions from
libraries, which are simply archives of ELF files, and performs the relocation
(see \S\ref{subsec:SignatureGenerator}) to merge the user's code and the library
functions code into a single executable.  In the latter, the linker does not use
any code from the libraries, but instead creates proxy/stub code which can
locate the entry point of each called library function at runtime.

Static linking, although it has drawbacks such as code duplication and is no longer
the default mode of linking on most platforms, is still used in cases where
dynamic linking is problematic. For example, unlike C, C++ and Fortran do not
have an agreed API and ABI (application binary interface), so not only object
files created by different C++/Fortran compilers can seldom be linked together,
object files created by different versions of the same compiler are not
guaranteed to interoperate either \cite{CppABI,GppABI}
For this reason, Fortran
compilers in particular, tend to use static linking. It is also not uncommon for
independent software vendors (ISVs) to ship only statically linked binaries to
avoid portability and library dependency issues.

On some platforms where the operating system is designed to be simple and
efficient, e.g. Cray XT's Catamount and IBM Blue Gene/L's Compute Node Kernel
(CNK), dynamic linking is usually not an option and static linking has to be
used \cite{Cray}.

A third case for static linking is the aforementioned compiler-specific code
snippets. They exist as object files or libraries and are almost always
statically linked.

For all of above reasons, library code snippets are the most important source
of signatures in our program build discovery tool.

\subsection{Symbol Versioning}
\label{subsec:SymbolVersioning}
Some dynamic libraries are self-annotated with version information in a uniform
format, and we use this information to identify both the library and its
version.

As mentioned in \S\ref{subsec:LibraryCodeSnippets}, dynamic linking has the
issue of interoperability. Historically, this was partly solved by having unique
file names for the dynamic libraries. The file names usually incorporate major
and minor release numbers, such as {\tt lib<name>.so.<major>.<minor>}.
The linker will then record the exact file names in the resulting
binaries' {\tt .dynamic} section. In 1995 Sun introduced a new and fine-grained
versioning mechanism in Solaris 2.5, which the GNU/Linux community soon adopted
\cite{SymVer}. In this scheme, each function name and symbol can be associated
with a version, and at the library level, a chain of version compatibility can be
specified. {\it The version of the library is then the highest version in the
version chain}.

As an example, in the GNU C runtime library ({\tt glibc}) source tree, one can find
version definition scripts containing the following

{\small
\begin{Verbatim}
libc {                libc {
  GLIBC_2.0 {           GLIBC_2.0
     malloc;            GLIBC_2.1
     free;              ...
     ...                GLIBC_2.10
  }                     ...
  ...                 }
  GLIBC_2.10 {
    malloc_info;
  }
}
\end{Verbatim}
}
The left-hand side specifies that {\tt malloc} and {\tt free} are versioned
GLIBC\_2.0 and {\tt malloc\_info} GLIBC\_2.10. The right-hand side indicates
GLIBC\_2.10 is compatible with GLIBC\_2.1, which is compatible with GLIBC\_2.0.
All of the versioning data are encoded in the {\tt .gnu.version\_d} section
({\tt d} for definition) of dynamic libraries when they are built. When a user
program is compiled and linked, a version-aware linker obtains versions of
called functions from the dynamic libraries and stores them in the resulting
binaries' {\tt .gnu.version\_r} section ({\tt r} for reference). At runtime, the
program loader-linker {\tt ld.so} first examines whether all version references
in the user's program binary can be satisfied or not, and determines to either abort
or continue.

Symbol versioning is used extensively in the GNU compiler collection (C,
C++, Fortran, and OpenMP runtime libraries), Myrinet MX/DAPL libraries, and
OpenFabrics/InfiniBand Verbs libraries. All of these instances adopt the same
version naming scheme: a unique label, e.g. {\tt GLIBC}, {\tt GLIBCXX}, or {\tt MX},
followed by an underscore and the version. Hence, our tool can recognize them
using a hard-coded list of labels and obtain their version by traversing
the version chain.

\subsection{Checksums}
Most dynamic libraries are less sophisticated and do not use symbol
versioning. Therefore, to recognize them, we resort to the traditional
approach of checksums. {\tt Md5sum} is a commonly used open-source utility
to produce and verify the MD5 checksum of a file, but it is file-structure
agnostic and fails to characterize ELF dynamic libraries on platforms (e.g.
Red Hat Enterprise Linux) where the prelinking/prebinding technology
\cite{Prelink} is used. Prelinking is intended to speed up the runtime loading
and linking of dynamic libraries when a program binary is launched. To
achieve this, a daemon process will periodically update the dynamic
libraries' relocation table. The side effect of prelinking
is MD5 checksum mismatch, as part of the file content has been changed. To
defeat this effect, we calculate the MD5 checksum
over the \textsec{} only for ELF files.

\section{Implementation}
\label{sec:Implementation}
Our implementation is based on the pattern matching engine of the
open-source anti-virus package ClamAV \cite{ClamAV}, with additional code
to support symbol versioning. The implementation comprises two tools: a
signature generator and a signature scanner. The signature generator parses
ELF files and outputs ClamAV-formatted signature files. The signature scanner
takes as input the signature files and the user's program binary and outputs
all possible matches. In the following, we discuss ClamAV's signature
formats and matching algorithms and
how we leverage ClamAV in our implementation.

\subsection{ClamAV Design}
ClamAV signatures can be classified as one of the following types, in the order
of increasing complexity and power: {\bf MD5}, {\bf basic}, {\bf regular
expression (regex)}, logical, and bytecode. Our implementation makes
use of the first three types because they can be generated automatically
(see \S\ref{subsec:SignatureGenerator}).

A basic signature is a hexadecimal string. ClamAV's scanning engine handles
this type of signature with a modified version of the classical
Boyer-Moore string searching algorithm called Wu-Manber.
A regex signature is a basic signature with wildcards. Our implementation
use two kinds of wildcards extensively: {\tt ??} (to match any byte) and
{\tt \{n\}} (to match any consecutive $n$ bytes). ClamAV's scanning engine
handles regex signatures with the Aho-Corasick (AC) string searching
algorithm, which can match multiple strings concurrently at the cost of
consuming more memory. The AC algorithm starts with a preprocessing
phase: Take a set of wildcard-free strings to create a finite automaton.
The scanning phase is simply a series of state transitions in this finite
automaton. ClamAV utilizes the AC algorithm as follows: Every regex
signature is broken into basic signatures (separated by wildcards), and a
single finite automaton (implemented as a two-level 256-way ``trie'' data
structure) is created from all of these basic signatures. If all
wildcard-free parts of a regex signature are matched, ClamAV checks
whether the order and the gaps between the parts satisfy the specified
wildcards.

For completeness we briefly mention the remaining two signature types. We
do not use them because we do not yet find automatic ways to create them.
Logical signatures allow combining of multiple regex signatures using
logical and arithmetic operators.
Bytecode signatures further extend logical signatures and offer the maximal
flexibility. Bytecode signatures are actually ClamAV plug-ins compiled from
C programs into LLVM bytecodes, and hence allow arbitrary algorithmic
detections of patterns.

\subsection{Signature Generator}
\label{subsec:SignatureGenerator}
For dynamic libraries ({\tt .so} files), the signature generator computes the MD5
checksums over their \textsecs{} and outputs the ClamAV-conformant
MD5 signature files.

Compiler-specific code snippets and static library code reside in ELF
{\tt .o} (object) and {\tt .a} (library archive) files. In the following
discussions we only focus on {\tt .o} file handling because an {\tt .a} file is
just an archive of multiple {\tt .o} files. Our signature generator
extracts \textsecs{} from {\tt .o} files, and outputs, for
each \textsec{}, a basic or regex signature of 16-255 bytes length (excluding
the wildcards.) We describe this process in depth as follows.

First, {\it a signature is not just bytes from the \textsec{} verbatim}.
When a source file is compiled into an {\tt .o} file, the addresses of
unresolved function names and symbols in this {\tt .o} file are unknown and
have to be left empty. It is during the linking phase that these addresses
are resolved and assigned by the linker. This process is called
{\it relocation} \cite{Linker}. To facilitate the relocation, the compiler emits
one relocation table for each \textsec{}. Each entry of a relocation table
specifies the symbol name to be resolved, the offset into the \textsec{}
which contains the address to be assigned, and the relocation type. When we
create a signature from the bytes of a \textsec{}, we have to {\it mask the
bytes which are reserved for addresses yet to be computed}. To illustrate,
suppose that we compile the following source code into an
{\tt .o} file:

{\small
\begin{Verbatim}
#include <stdlib.h>
void foo() {
  char *buf = malloc(10);
}
\end{Verbatim}
}

On x86, the disassembly of the generated {\tt .o} file would be
(using the GNU {\tt objdump} utility):

{\small
\begin{Verbatim}[commandchars=\\\{\}]
000000 <foo>:
 0: 55              push   %rbp
 1: 48 89 e5        mov    %rsp,%rbp
 4: 48 83 ec 10     sub    $0x10,%rsp
 8: bf 0a 00 00 00  mov    $0xa,%edi
 d: e8\fbox{00 00 00 00} callq  12 <foo+0x12>
12: 48 89 45 f8     mov    %rax,-0x8(%rbp)
16: c9              leaveq
17: c3              retq
\end{Verbatim}
}
and the corresponding relocation table is:
{\small
\begin{Verbatim}
OFFSET  TYPE          VALUE
00000e  R_X86_64_PC32 malloc+0xffffffff
                               fffffffc
\end{Verbatim}
}

Together, the above examples illustrate that the target of the {\tt callq} instruction
should be the address of a function named "malloc", and the address
should fill the 4 bytes (as specified by the {\tt R\_X86\_64\_PC32} relocation type)
starting at offset {\tt 0xe} (the boxed {\tt 00}'s). So if {\tt foo}, as a
library function, is used to create a user program binary, the linker
will take the byte stream {\tt 55 48 89 e5} \ldots {\tt c9 c3} and fill the bytes at
offset {\tt 0xe} through {\tt 0xe}+3 with the actual address of malloc.
Thus, to identify {\tt foo}, we create a ClamAV regex signature as:
{\small
\begin{verbatim}
55 48 89 e5 48 83 ec 10 bf 0a 00 00
00 e8 ?? ?? ?? ?? 48 89 45 f8 c9 c3
\end{verbatim}
}

The second consideration is the signature size. As will be seen in
\S\ref{subsec:LibraryIdentification} a \textsec{} can be as
big as four megabytes. Using the entire \textsec{} could lead to long preprocessing time
and large disk/memory storage space. Therefore, we impose
an upper limit on the signature size to be 255 bytes. We think 255
is a reasonable size, as there are $256^{255}$ possible distinct 256-byte streams,
which is large enough to have few collisions/false positives. For a
\textsec{} of $n> 256$ bytes, we use the tailing 255/3=85 bytes
$x_1x_2 \ldots x_{85}$ of the first third portion, the tailing 85 bytes
$y_1y_2 \ldots y_{85}$ of the middle third, and the
tailing 85 bytes $z_1z_2 \ldots z_{85}$ of the last middle third,
and form a regex signature as:
\[
x_1 x_2 \ldots x_{85} \; \{l\} \; y_1 y_2 \ldots y_{85} \; \{m\} \; z_1 z_2 \ldots z_{85}
\]
where $l = \lfloor n/3 \rfloor -85$ and $m=l+(n \% 3)$. We also ignore \textsecs{}
which are shorter than 16 bytes. This cut-off is chosen because the size
of an x86 instruction varies between 1 and 16 bytes, and since we do not
decode the bytes back to x86 instructions, we do not know the instruction boundaries
and have to make a conservative assumption. Besides, signatures that
are too short could result in many false positives.

The third consideration is an {\tt .o}  file could contain more than one \textsec{}.
This happens in GNU Fortran's static library, which is created with the
{\tt -ffunction-sections} compiler flag. This flag instructs the compiler to
put each function in its own \textsec{} instead of all functions from the same
source file in one single \textsec{}. So for a Fortran function, say {\tt foo},
the compiler creates a section named {\tt .text.foo} which consists of {\tt foo}'s
code only \footnote{This optimization reduces the size of statically linked program
binaries because it eliminates dead code, i.e. functions which are unused
but included nevertheless because they are in the same source files as the
used functions.}. In such a situation, our tool emits one signature for one
such \textsec{}.

\subsection{Signature Scanner}
\label{subsec:SignatureScanner}
The signature database is organized as a collection of signature files, each of
which contain signatures from a specific compiler/library, e.g. Intel Fortran
compiler, Intel MKL, MVAPICH, etc. Each signature file is annotated manually to
indicate the package name and version. The scanner takes as input this database
and the user's program binary and outputs all possible matches. For dynamic
library identification, it uses the {\tt ldd} command to obtain the library
pathnames. It then extracts their symbol versioning data  (if there is any) and
compares against a list of known labels, as explained in
\S\ref{subsec:SymbolVersioning}. For those without symbol versioning, the
scanner checks their MD5 checksums against those in the database.

For compiler and static library identification, the scanner loads the program
binary's {\tt .text} and {\tt .comment} sections (compiler meta-data are treated
as basic signatures) and runs them through the ClamAV matching engine.
By default ClamAV stops as soon as it spots a match, so to find all matches,
we modify it by repeatedly zeroing out the matched area and rerunning the
engine, until no match can be found.

\section{Evaluation}
\label{sec:Evaluation}
We evaluate our approach with both toy programs and real-world HPC software
packages from two HPC sites. We compile toy programs with
a variety of compilers to test the effectiveness of source compiler
identification. We use the existing HPC software packages to assess not only
the compiler and library recognition but also ClamAV's scanning
performance.

\subsection{Compiler Identification}
\label{subsec:CompilerIdentification}
We examine fourteen compilers on the x86-64 Linux platform and we summarize our
findings in Table~\ref{tbl:CompilerIdentification}. We locate the
compiler-specific code snippets by enabling the verbosity flag in building the toy
programs. This flag is supported by all compilers and it can display exactly
where and which {\tt .a} and {\tt .o} files are used in the compilation process.
The toy programs we constructed, e.g. "Hello, World'' and matrix multiplication,
are short and use only basic language features and APIs, so they can highlight
the usefulness of our approach. All test cases are compiled with each
compilers' default settings.

\begin{table}

\footnotesize{

\begin{tabularx}{\linewidth}{|>{\hsize=1\hsize}X|>{\hsize=0.7\hsize}X|>{\hsize=1\hsize}X|>{\hsize=0.7\hsize}X|>{\hsize=1.6\hsize}X|}
\hline
Compiler & Note & Version & Meta Data & Code Snippet Source \\
\hline\hline
Absoft   & F,O  & 11.1    &      & liba*.a   \\
\hline
Clang    & C,L  & 2.8     &      &           \\
\hline
Cray     &      & 7.1, 7.2  & V    & libcsup.a, libf*.a, libcray*.a \\
\hline
G95      & F,G  & 0.93    & V    & libf95.a \\
\hline
GNU      & G    & 4.1, 4.4, 4.5 & V    & crt*.o, libgcc*.a \\
\hline
Intel    &      & 9.x thru 12.0  & I    & libirc*.a, libfcore*.a \\
\hline
Lahey-Fujitsu  & F & 8.1  & I    & fj*.o, libfj*.a \\
\hline
LLVM-GCC & G,L   & 2.8    & V    & \\
\hline
NAG      & F,$\dagger$     & 5.2    &      & libf*.a \\
\hline
Open64   & O,$\ddagger$    & 4.2  & V    & libopen64*.a, libf*.a \\
\hline
PathScale & O,$\ddagger$   & 3.2, 4.0 & V  & lib*crt.a, libpath*.a \\
\hline
PCC      & C     & 0.99     & V  & crt*.o, libpcc*.a \\
\hline
PGI      &       & 6.x thru 11.x & V   & libpgc.a, libpgf*.a, f90*.o, pgf*.o \\
\hline
Sun Studio &     & 12.x     & V  & crt*.o, libc\_supp.a, libf*.a \\
\hline
\end{tabularx}
}
\caption{Compiler identification.
 C: C/C++ compiler only. F: Fortran compiler only. G: uses GNU codebase. I: has unique
meta data. L: uses LLVM codebase. O: uses Open64 codebase. V: meta data have both
brand string and version number.
$\dagger$: is actually a Fortran-to-C
converter with GCC as backend.
$\ddagger$: inserts FTZ/DAZ-enabling prolog code (see \S\ref{subsec:CompilerSpecificCodeSnippets})
but this code is not in any {\tt .a}/{\tt .o} files so
we manually produce its signature.
}
\label{tbl:CompilerIdentification}
\end{table}

\begin{table}
\footnotesize{
\begin{tabularx}{\linewidth}{|>{\hsize=1\hsize}X|>{\hsize=1\hsize}X|>{\hsize=1\hsize}X|>{\hsize=1\hsize}X|}
\hline
Library     & Version (Compiler)  & Code Snippet Source & Mean and StdDev {\tt .text} size in KB \\
\hline\hline
ACML        & 4.4.0 (I,P)     & libacml*.a    & 11.1, 70.8   \\
\hline
Cray LibSci & 10.4.0 (G,I,P)  & libsci*.a     & 3.4, 4.9     \\
\hline
Intel MKL   & 8.0, 8.1, 9.1   & libmkl*.a     & 4.6, 9.0  \\
\cline{2-4}
            & 10.x            & libmkl\_core.a & 4.2, 16.6 \\
\hline
Cray MPI    & 3.5.1 (G,I,P)   & libmpich*.a   & 1.3, 2.6  \\
\hline
MPICH       & 1.2.7mx (G,I)   & libmpich.a    & 1.2, 2.7  \\
\hline
MVAPICH2    & 1.4, 1.5 (I)    & libmpich.a    & 2.6, 4.8  \\
\hline
\end{tabularx}
}
\caption{Library identification.
G: GNU. I: Intel. P: PGI.
}
\label{tbl:LibraryIdentification}
\end{table}

As an example, the "Hello, World'' program compiled with Intel
compiler 12.0 yields the following output from our scanner. It
gives the number of matches and total size of matches against each
signature file:
{\footnotesize
\begin{verbatim}
(3 times, 6992 bytes) Intel Compiler Suite 12.0
(2 times, 200 bytes)  GCC 4.4.3
\end{verbatim}
}

We have the following observations. 1. Many compilers strive to be compatible
with the GNU development tools and runtime environment, so they also use GNU's
code snippets. Therefore, GCC becomes a common denominator and is ubiquitous in
the scanning results. The above output is typical: The Intel compiler locates
the system's default GCC installation (version 4.4.3 in this case) and uses its
{\tt crtbegin.o} and {\tt crtend.o} in the compilation. These two {\tt .o} files
handle the {\tt .ctors} section as discussed in
\S\ref{subsec:CompilerSpecificCodeSnippets}.

2. As opposed to C, the Fortran compiler space is very fragmented, with each
compiler having its own implementation of language intrinsics and extensions.
Hence, we can spot a Fortran compiler by just examining the runtime library code
used. The same is true for OpenMP and UPC.

3. It is possible to recognize different versions of the same compiler. To
demonstrate, we wrote a simple Fortran program which calls the {\tt matmul}
intrinsic to perform matrix multiplications and compiled it with PGI 11.0.
The result is as follows:
{\footnotesize
\begin{verbatim}
(58 times, 346766 bytes) PGI Fortran Compiler 11.x
(48 times, 56833 bytes)  PGI Fortran Compiler 8.x
(45 times, 118288 bytes) PGI Fortran Compiler 10.x
(42 times, 49895 bytes)  PGI Fortran Compiler 7.x
(32 times, 82808 bytes)  PGI Compiler Suite 11.x
(29 times, 57166 bytes)  PGI Compiler Suite 7.x
....
(2 times, 200 bytes) GCC 4.4.3
\end{verbatim}
}
The matches include both the Fortran runtime library and compiler-specific code
snippets, which are shared by C/C++ and Fortran compilers. The result also
implies that PGI reuses a significant amount of code across each release.
We scrutinized the code snippets which matched both versions 7.x and 11.x
and found their functionality includes memory operations (allocate, copy,
zero, set), I/O setup (open, close), command-line argc/argv handling, etc.

4. Compilers which share codebase are not easily distinguishable. Examples
include Open64 and PathScale, GNU and LLVM-GCC, etc. In these cases, only
the compiler-specific meta data can tell them apart, and Clang is thus
far the only compiler which defies our inference efforts.

\subsection{Library Identification}
\label{subsec:LibraryIdentification}
We applied the scanner to a subset of HPC applications (Amber~\cite{Amber},
Charmm~\cite{Charmm}, CPMD~\cite{CPMD}, GAMESS~\cite{GAMESS},
Lammps~\cite{Lammps}, NAMD~\cite{NAMD}, NWChem~\cite{NWChem},
PWscf~\cite{PWscf}) from two HPC sites (a 3456-core Intel-based commodity
PC cluster at our center and a 672-core Cray XT5m at Indiana University).
We gathered signatures from numerical and MPI libraries which
we know have been linked statically in the application builds. The
libraries and the size of their constituent {\tt .o} files are summarized
in Table~\ref{tbl:LibraryIdentification}. Numerical libraries tend to have more
{\tt .o} files and larger code size per {\tt .o} file. The explanation
is various processor-specialization codes and aggressive loop unrolling. For
example, ACML 4.4.0-ifort64's {\tt libacml.a} has 4.5K {\tt .o} files, with
the largest (4.1 MB code size) being an AMD-K8-tuned complex matrix multiplication
(zgemm) kernel, and Intel MKL 10.3.1's {\tt libmkl\_core.a} has 44K
{\tt .o}'s, with the largest (1.4 MB) being an Intel-Nehalem-optimized
batched forward discrete Fourier transform code.

For the test we create a signature database exclusively from the
aforementioned libraries. It has 100K signatures  and the predominant
signature type is regex. The 21 HPC application binaries under test have a
mean code size of 13.3 MB and the largest is NWChem 6.0 on Cray (39.4 MB,
mainly due to static linking, as in \S\ref{subsec:LibraryCodeSnippets}).
We build the (single-threaded) scanner with Intel compiler 12.0 and we run
the scan on a 2.5 GHz Intel Xeon L5420 "Harpertown'' node and a 2.8
GHz X5560 "Nehalem'' node. The results show that the scanner can correctly
identify all used libraries. The scanning time $t$ (in seconds) can be best
described by the linear regressions $t=-1.11 + 7.23x$ (Harpertown) and
$t=-5.44 + 6.98x$ (Nehalem) where $x$ is the code size in MB, and the peak
memory usage is 195 MB.

\section{Discussion}
\label{sec:Discussion}
Our methodology of identifying the source compiler depends on the
idiosyncrasies of the x86 platform and compilers. We also explored the two
major compilers, GCC and IBM XL, on the PowerPC platform, and did not find
discernible compiler-specific code snippets. IBM XL compilers do inscribe
their brand strings in the \comsec{}, but in general, content in
\comsec{} is subject to tampering. For example, the following line in a C
program:
{\small
\begin{Verbatim}
__asm__(".ident \"foo\"");
\end{Verbatim}
}
will emit ``{\tt foo}'' to the \comsec{}. This makes \comsec{}
a less reliable source of compiler provenance from a general
perspective of software forensics.

Another issue is that a compiler inserts its characteristic prolog code only
when it is compiling the source file which contains the {\tt main}
function. So if different source files are compiled with different
compilers, the resulting program binary could lack the compiler-specific
code snippets one would expect. In addition, in Intel compiler's case, it
does not insert processor-dispatch code if the optimization is turned off
either explicitly (with {\tt -O0}) or implicitly (e.g. with {\tt -g}).

Our approach cannot discover the compilation flags used in the program
build process. Some compilers offer a switch to record the command-line
options inside either {\tt .comment} or other sections. For example, Intel
has {\tt -sox},  GCC has {\tt -frecord-gcc-switches} (recorded in
{\tt .GCC.command.line} section), and Open64/PathScale and Absoft do it by
default. We expect this self-annotation feature to be more widely embraced
by compiler developers, as they move toward better compatibility with GCC,
and used by HPC programmers, as it greatly aids debugging and performance
analysis.

\section{Related Work}
\label{sec:RelatedWork}
ALTD~\cite{ALTD} is an effort to track software and library usage at HPC
sites. It takes a proactive approach by intercepting and recording every
invocation of the linker and the job scheduler. Our work is complementary
in that it performs post-mortem analysis and works on systems without ALTD.

The work by Rosenblum {\em et al}~\cite{Rosenblum} is the first attempt to
infer the compiler provenance. They used sophisticated machine learning by
modeling and classifying the code byte stream as a linear chain Conditional
Random Field. As in most supervised learning systems, a lengthy training
phase is required.
The resulting system can then infer the source compiler with a probability.
Their approach has several drawbacks, which our method addresses: They
focus solely on executable code and ignores other parts of ELF files, the
preprocessing/training phase, albeit one-time, is slow and complex, the
model parameters cannot be updated incrementally with ease when a new
compiler is added, and it is unclear if their model can discern the nuances
among different versions of the same compiler.

Kim's approach~\cite{Kim} is closest to ours in spirit, but it misses the
key feature in our implementation: the relocation table. It produces a signature by
copying the first 25 bytes of a library function code {\it verbatim}. With
such a short signature and lack of relocation information, his tool has
very limited success in identifying library code snippets.

\section{Conclusion}
\label{sec:Conclusion}
Compilers and libraries provenance reporting is crucial in an auditing and
benchmarking framework for HPC systems. In this paper we present a simple and
effective way to mine this information via signature matching. We also
demonstrate that building and updating a signature database is straightforward
and needs no expert knowledge. Finally, our tests show excellent scanning speed
even on very large program binaries.

\section*{Acknowledgments}
This work is supported by the National Science Foundation under award
number OCI 1025159. We would like to thank Gregor von Laszewski for providing
access to FutureGrid computing resources.

\bibliographystyle{ieeetr}
\footnotesize{
\bibliography{appsig}

\begin{thebibliography}{10}

\bibitem{XDMoD}
T.~R.~Furlani {\em et~al.}, ``Performance metrics and auditing framework using
applications kernels for high performance computer systems.'' In preparation.

\bibitem{SPEC}
http://www.spec.org

\bibitem{Modules}
http://modules.sf.net

\bibitem{SoftEnv}
http://www.teragrid.org/userinfo/softenv/

\bibitem{ELF}
M.~Wilding and D.~Behman, ``Self-service {Linux}: Mastering the art of problem 
determination.'' Prentice Hall, 2005.

\bibitem{ELF2}
``{System V} application binary interface - {AMD64} architecture processor 
supplement.'' http://www.x86-64.org/documentation/


\bibitem{Agner}
A.~Fog, Chapter 13 of ``Optimizing software in C++: An optimization guide for Windows, Linux and Mac platforms.''
  http://www.agner.org/optimize/
  
\bibitem{Denormal}
I.~Dooley and L.~Kale, ``Quantifying the interference caused by subnormal floating-point values.''
  {\em The Workshop on Operating System Interference in High Performance Applications (OSIHPA)}, 2005.

\bibitem{CppStandard}
\S8.5 of ``Working Draft of Standard for Programming Language C++, Document No. N1905.''
http://www.open-std.org
  
\bibitem{Linker}
J.~R.~Levine, ``Linkers and loaders.'' Morgan Kaufmann, 1999.

\bibitem{ClamAV}
T.~Kojm, http://www.clamav.net

\bibitem{SymVer}
D.~J.~Brown and K.~Runge, ``Library interface versioning in {Solaris} and {Linux}.''
  {\em The 4th Annual Linux Showcase (ALS) {\&} Conference}, 2000.

\bibitem{ALTD}
B.~Hadri, M.~Fahey, and N.~Jones, ``Identifying software usage at {HPC} centers 
  with the automatic library tracking database.'' {\em Proceedings of the 2010
  TeraGrid Conference}.

\bibitem{CppABI}
N.~Sidwell, ``A common vendor {ABI} for {C++} -- {GCC}'s why, what and not.'' 
  {\em Proceedings of the 2003 ACCU Conference}.

\bibitem{GppABI}
http://gcc.gnu.org/onlinedocs/libstdc++/manual/abi.html

\bibitem{Rosenblum}
N.~Rosenblum, B.~Miller, and X.~Zhu, ``Extracting compiler provenance from 
  program binaries.'' {\em The workshop on Program Analysis 
  for Software Tools and Engineering (PASTE)}, 2010.

\bibitem{Cray}
G.~Johansen and B.~Mauzy, ``{Cray XT} programming environment's implementation 
  of dynamic shared libraries.'' {\em Cray User Group (CUG) Conference}, 2009.

\bibitem{Prelink}
J.~Jelinek, http://people.redhat.com/jakub/prelink.pdf


\bibitem{Kim}
J.~S.~Kim, ``Recovering debugging symbols from stripped static compiled binaries.''
  {\em Hakin9 Magazine}, June 2009. http://0xbeefc0de.org/papers/

\bibitem{Amber}
D.~A.~Case {\em et~al.}, ``The Amber biomolecular simulation programs.'' 
{\em J. Comp. Chem.} v 26, 1668-1688 (2005). 

\bibitem{Charmm}
B.~R.~Brooks {\em et~al.}, ``CHARMM: The biomolecular simulation program.''
{\em J. Comp. Chem.} v 30, 1545-1615 (2009).

\bibitem{CPMD}
http://www.cpmd.org

\bibitem{GAMESS}
M.~W.~Schmidt {\em et~al.}, ``General atomic and molecular electronic structure system.''
{\em J. Comp. Chem.} v 14, 1347-1363 (1993).

\bibitem{Lammps}
S.~J.~Plimpton, ``Fast parallel algorithms for short-range molecular dynamics.''
{\em J. Comp. Phys.} v 117, 1-19 (1995). 

\bibitem{NAMD}
J.~C.~Phillips {\em et~al.}, ``Scalable molecular dynamics with NAMD.''
{\em J. Comp. Chem.} v 26, 1781-1802 (2005).

\bibitem{NWChem}
M.~Valiev {\em et~al.}, ``NWChem: a comprehensive and scalable open-source solution for large scale molecular simulations.''
{\em Comput. Phys. Commun.} v 181, 1477 (2010).

\bibitem{PWscf}
P.~Giannozzi {\em et~al.}, http://www.quantum-espresso.org

\end{thebibliography}
}
\end{document}